\documentclass[a4paper,11pt]{article}

\usepackage{pos}
\usepackage{wrapfig}

\title{Cosmic Ray Anisotropy with 11 Years of IceCube Data}

\ShortTitle{11 Years of Anisotropy in IceCube}

\author{The IceCube Collaboration \\{\normalsize \normalfont(a complete list of authors can be found at the end of the proceedings)}\\}

\emailAdd{mcnally\_ft@mercer.edu}

\abstract{

The IceCube Observatory provides our highest-statistics picture of the cosmic-ray arrival directions in the Southern Hemisphere, with over 700 billion cosmic-ray-induced muon events collected between May 2011 and May 2022. Using the larger data volume, we find an improved significance of the PeV cosmic ray anisotropy down to scales of $6^\circ$. In addition, we observe a variation in the angular power spectrum as a function of energy, hinting at a relative decrease in large-scale features above 100\,TeV. The data-taking period covers a complete solar cycle, providing new insight into the time variability of the signal. We present preliminary results using this up-to-date event sample.
\vspace{4mm}
{\bfseries Corresponding authors:}
Frank McNally$^{1*}$, Rasha Abbasi$^{2}$, Paolo Desiati$^{3}$, Juan Carlos D\'iaz V\'elez$^{3}$, Christina Cochling$^{1}$, Katherine Gruchot$^{2}$, William Hayes$^{2}$, Andrew Moy$^{2}$, Emily Schmidt$^{1}$, and Andrew Thorpe$^{1}$\\
{$^{1}$ \itshape Mercer University, U.S.A.}\\
{$^{2}$ \itshape Loyola University Chicago, U.S.A.}\\
{$^{3}$ \itshape University of Wisconsin - Madison, U.S.A.}\\[4mm]
$^*$ Presenter

\ConferenceLogo{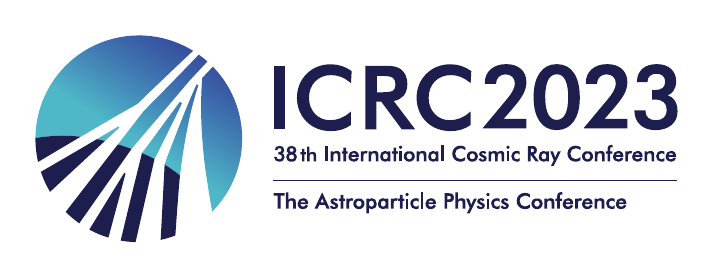}

\FullConference{The 38th International Cosmic Ray Conference (ICRC2023)\\ 26 July -- 3 August, 2023\\ Nagoya, Japan}
}

\begin{document}

\maketitle

%\linenumbers

\section{Introduction}\label{sec1}

%Astrophysics observations aim to identify the sources of the cosmic rays, the acceleration mechanisms that inject them into space, and their propagation properties through magnetized plasmas. Gamma-ray and neutrino experiments are dedicated to pinpointing the remote location of the cosmic-ray sources in the Milky Way or other galaxies. The recent detection of high-energy galactic gamma-ray galactic sources might already provide hints~\citep{pevatron-tibet, pevatron-lhaaso}. On the other hand, cosmic ray particles observed on Earth carry with them the properties of the medium traversed in their journey. The study of cosmic ray transport is the key to understanding their astrophysical origin. Most investigations rely on the information derived from energy spectrum and composition measurements. The arrival direction distribution, however, offers a method to study the cosmic ray pitch angle distribution in the interstellar medium, which provides a direct probe of interstellar diffusion properties~\citep{snowmass2021}.

One of the goals of astrophysical observations is to identify the sources of cosmic rays and the mechanism by which they are accelerated and injected into the interstellar medium. The cosmic ray particles observed on Earth carry the general properties of the sources from which they originated and the medium traversed in their journey. The study of cosmic ray transport is a key to understanding their astrophysical origin. Most investigations rely on the information derived from energy spectrum and mass composition measurements. However, measuring their arrival direction offers a method to study the cosmic ray pitch angle distribution in the interstellar medium, which provides a direct probe of interstellar diffusion properties~\citep{snowmass2021}.
Several theoretical models predict an anisotropy in the cosmic ray arrival directions that results from the distribution of sources in the Milky Way and diffusive propagation of these particles~\citep{erlykin_2006, Blasi:2012jan, Ptuskin:2012dec, Pohl:2013mar, Sveshnikova:2013dec, Kumar:2014apr, Mertsch:2015jan, ahlers2014, Ahlers:2015dwa, ahlers_2016}. 

Over the last two decades, large and modern ground-based experiments have provided high-volume data that facilitate state-of-the-art analyses of cosmic ray anisotropy in the TeV-PeV energy range (see~\citep{abbasi_2010, abbasi_2011, IceCube:2012feb, Aartsen:2012ma, aartsen_2016, DiazVelez2017, lhaaso_2021} and references therein).
%~\citep{nagashima_1998, hall_1999, Amenomori_2005, Amenomori_2006, Takita:2007hja, guillian_2007, Abdo_2008, Abdo_2009, Aglietta_2009, Iuppa:2010zf, munakata_2010, TIBETASg:2011dha, deJong:2012hk, Cui:2011sda, bartoli_2013, abeysekara_2014, Bartoli_2015, Amenomori:2017jbv, Bartoli_2018, abeysekara_2018, abbasi_2010, abbasi_2011, IceCube:2012feb, aartsen_2013, aartsen_2016, Bourbeau:2017Fy, DiazVelez2017, fermi_2019, ahlers_2019, grapes3_2021, lhaaso_2021}. 
%
% Problem children: grapes3_2021, ahlers_2019
% ICRC 2007: amenomori_2007 --> Takita:2007hja
% ICRC 2009: zhang_2009 --> Iuppa:2010zf
% ICRC 2011: shuwang_2011 --> Cui:2011sda, dejong_2011 --> deJong:2012hk, amenomori_2011 --> TIBETASg:2011dha
% ICRC 2017: bourbeau_2017 --> Bourbeau:2017Fy
The observed anisotropy (up to the order of 10$^{-3}$ in relative intensity) possesses a complex structure manifesting at different angular scales. It also appears to evolve with energy as well. New generation experiments currently under design (e.g., SWGO~\citep{swgo_whitepaper, swgo_astro2020} and IceCube-Gen2~\citep{icecube-gen2}) will further enhance the scientific reach, in addition to making it possible to provide additional full-sky coverage for unbiased anisotropy observations.

This work presents the up-to-date cosmic ray anisotropy results observed by the IceCube Observatory using about 700 billion events collected from 2011 to 2022. IceCube is an astroparticle observatory at the geographic South Pole, consisting of a cubic kilometer of instrumented volume over a kilometer under the ice surface, sensitive to cosmic rays via the muonic component of air showers~\citep{Abbasi:2008ym}. The observation shows the evolution of the arrival direction distribution from 10\,TeV to 530\,TeV median energy, with high statistical significance.

\section{Data Analysis}

% Overview paragraph. Introduction to the data. Connection to previous analysis. What's changed?
The methods used in this analysis are largely consistent with previous anisotropy studies in IceCube~\citep{abbasi_2010, abbasi_2011, IceCube:2012feb, aartsen_2016, Bourbeau:2017Fy}. We incorporated improvements in the cosmic-ray shower physics, the experiment's response in our Monte Carlo simulation, and systematics since the last major publication, as presented in Ref.~\citep{Abbasi:2021kjd}. The most significant change lies in the increased statistics (over a factor of two compared to the results in Ref.~\citep{aartsen_2016}), allowing for more detailed energetic studies and spanning a full solar cycle.

% How do we make our maps (part 1)}
To analyze cosmic-ray anisotropy, we first need to build a reference map representing the detector's sensitivity to an isotropic cosmic-ray flux. For this, we use a time scrambling method~\cite{abbasi_2011}. The data map contains the directions of events in sidereal coordinates (RA, Dec) calculated from their local coordinates ($\theta, \phi$) and arrival time $t$. To build the reference map, each event in the dataset is assigned a random time sampled from the time distribution of all events over a time window $\Delta t$. This approach maintains both the local arrival direction distribution and the temporal distribution of events while scrambling their sidereal coordinates. For every real event in the data, we generate 20 time-scrambled events in order to reduce statistical fluctuations. 
% NOTE for v6: removing the sentence below along w/ the equation
%The ratio of data to background events is defined as $\alpha$ for significance calculations (see Eq.~\ref{eq:risigma}). 
The IceCube detector rate is very stable over a period of 24\,h, which means we can choose $\Delta t$ = 24\,h. 

% How do we make our maps (part 2)
Once we have constructed a reference map, we can compute the relative intensity for each pixel in the map by 
\begin{equation}
    \frac{\delta I_i}{I_i} = \frac{N_i- \langle N_i \rangle}{\langle N_i \rangle}~,
    \label {eq:ri}
\end{equation}
where $N_i$ and $\langle N_i \rangle$ are the number of events at the $i^\mathrm{th}$ pixel of the data map and the reference map, respectively.
The statistical significance of any deviation from the reference level can be calculated using the Li \& Ma method~\citep{LiMa1983}. 
% NOTE for v6: removing equation
%The significance $s$ of pixel $i$ is given by
%
%\begin{equation}
%    s_i = \sqrt{2} \left\{ N_i \log \Bigg[ \frac{1 + \alpha}{\alpha} \Big( \frac{N_i}{N_i + N_o} \Big) \Bigg] + N_o \log \Bigg[ (1+\alpha) \Big( \frac{N_o}{N_i + N_o} \Big) \Bigg] \right\}^{1/2}~,
%    \label{eq:risigma}
%\end{equation}
%with $N_o = \langle N \rangle_i/\alpha$.
This method takes into account statistical fluctuations in data and reference counts.
%
%
%\begin{wrapfigure}{R}{0.5\textwidth}
%  \begin{center}
%    \includegraphics[width=0.48\textwidth]{ICRCtemplate/figures/IC86_Median_Energy_Spline.png}
%  \end{center}
%  \caption{Median cosmic-ray primary energy as a function of reconstructed shower direction ($\cos\theta$) and number of DOMs triggered ($\log_{10}N$). This table is the result of a 2D-histogram populated using simulated events, then smoothed with splines to avoid low-statistics artifacts at the highest energies.}
%  \label{fig:med_energy}
%\end{wrapfigure}
%
%
We apply ``top-hat" smoothing to the data and reference maps in order to increase the sensitivity to regions with excess at different angular sizes by adding all counts within a given angular radius of each pixel. 

%% NOTE figure for v4: including true energy distribution
%\begin{figure*}[ht]
%  \centering
%  \includegraphics[width=0.49\textwidth]{ICRCtemplate/figures/IC86_Median_Energy_Spline.png}
%  \includegraphics[width=0.49\textwidth]{ICRCtemplate/figures/IC86_Energy_Distributions_1.png}
%  \caption{\textit{Left:} Median cosmic-ray primary energy as a function of reconstructed shower direction ($\cos\theta$) and number of DOMs triggered ($\log_{10}N$). This table is the result of a 2D-histogram populated using simulated events, then smoothed with splines to avoid low-statistics artifacts at the highest energies. \textit{Right:} Normalized true energy distributions for low- and high-energy skymaps used to study large- and small-scale structure as a function of energy. The low-energy bin consists of events with $\log_{10}(E_\mathrm{reco}/\mathrm{GeV}) < 4.25$; the high-energy bin consists of events with $\log_{10}(E_\mathrm{reco}/\mathrm{GeV}) > 5.5$. The distributions feature a 25\% overlap.}
%  \label{fig:energy_sim}
%\end{figure*}

\begin{wrapfigure}{R}{0.5\textwidth}
  \begin{center}
    \includegraphics[width=0.48\textwidth]{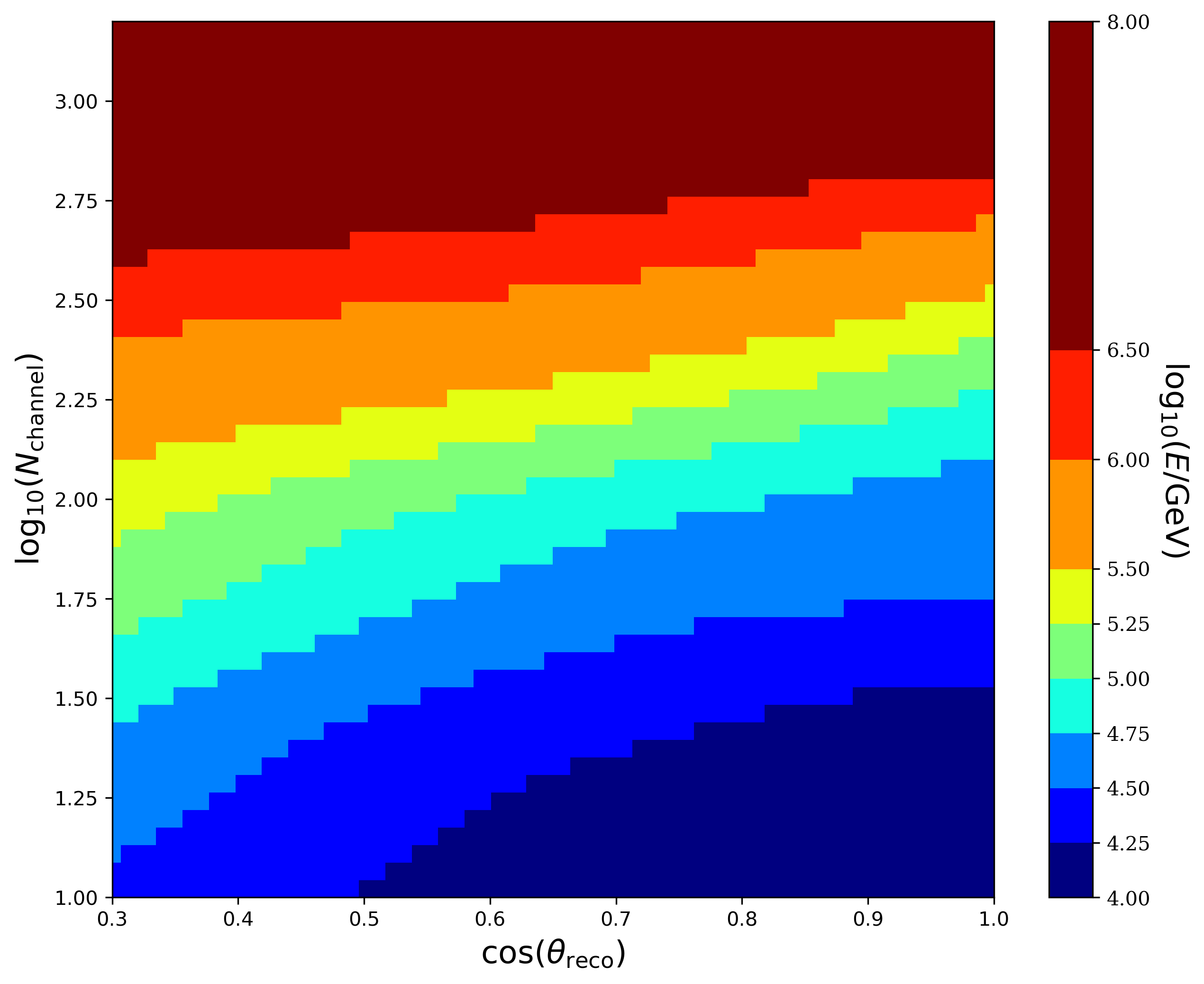}
  \end{center}
  \caption{Median cosmic-ray primary energy as a function of reconstructed shower direction ($\cos\theta$) and number of DOMs triggered ($\log_{10}N$). This table is the result of a 2D-histogram populated using simulated events, then smoothed with splines to avoid low-statistics artifacts at the highest energies.}
  \label{fig:energy_sim}
\end{wrapfigure}

% Energy reconstruction
The energy reconstruction is performed using a simple lookup table. Simulated events are weighted to a Gaisser H3a composition model~\cite{Gaisser:2012zz} and binned based on the number of digital optical modules (DOMs) triggered ($N_\mathrm{channel}$) and the reconstructed arrival direction of the cosmic ray primary ($\theta_\mathrm{reco}$). Both of these parameters correlate strongly with the energy of observed events --- more energetic showers trigger more DOMs on average, and showers with larger zenith angles must pass through more of the atmosphere, filtering out lower-energy events. Each bin is then assigned its median energy value and we smooth the table to avoid statistical artifacts at the highest energies where simulation is limited (Fig.~\ref{fig:energy_sim}). Using this table, each event is assigned an energy bin based on its $N_\mathrm{channel}$ and $\theta_\mathrm{reco}$ values. 

% Motivation for split large- and small-scale structure
% Note: organization confusion. Should this be in Results?
The reconstructed energy bins have true energy distributions that overlap significantly, but we can still use them to study changes in the observed anisotropy as a function of energy. Previous studies show wide regions of excess and deficit (right ascensions of 30-120$^\circ$ and 150-250$^\circ$, respectively) at low energies, replaced by a wide, dominant deficit centered at a right ascension of about $80^\circ$~\citep{aartsen_2016, Abbasi:2021kjd}. Despite reduced statistics, the high-energy deficit is sufficiently strong to make a visible impact on the large-scale structure map, previously reported using all observed events. For this reason, we now show the large- and small-scale structure visible in the Southern Hemisphere using only the lowest energy bin. This approach should help disentangle the separate low- and high-energy signals. 
% NOTE for v4: original sentence below
%A high-energy map is also shown, consisting of the three highest energy bins in our analysis (i.e., events with $\log_{10}(E/\mathrm{GeV}) > 5.5$) --- the true energy distributions of the two skymaps are shown in Fig.~\ref{fig:energy_sim} (\textit{right}).
%
A high-energy map is also shown, consisting of the three highest energy bins in our analysis (i.e., events with $\log_{10}(E/\mathrm{GeV}) > 5.5$). The true energy distributions have median energies of 13 and 530\,TeV, respectively, with a 25\% overlap.

\section {Results}

\begin{figure*}[ht]
  \centering
  \includegraphics[width=0.49\textwidth]{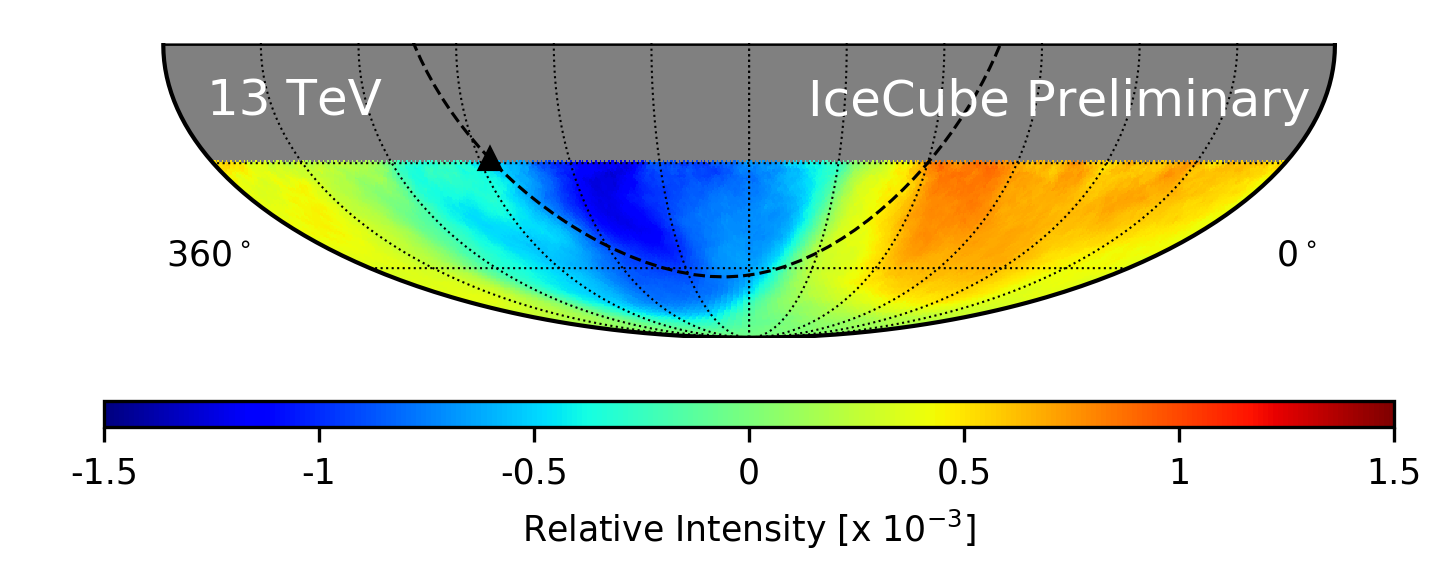}
  \includegraphics[width=0.49\textwidth]{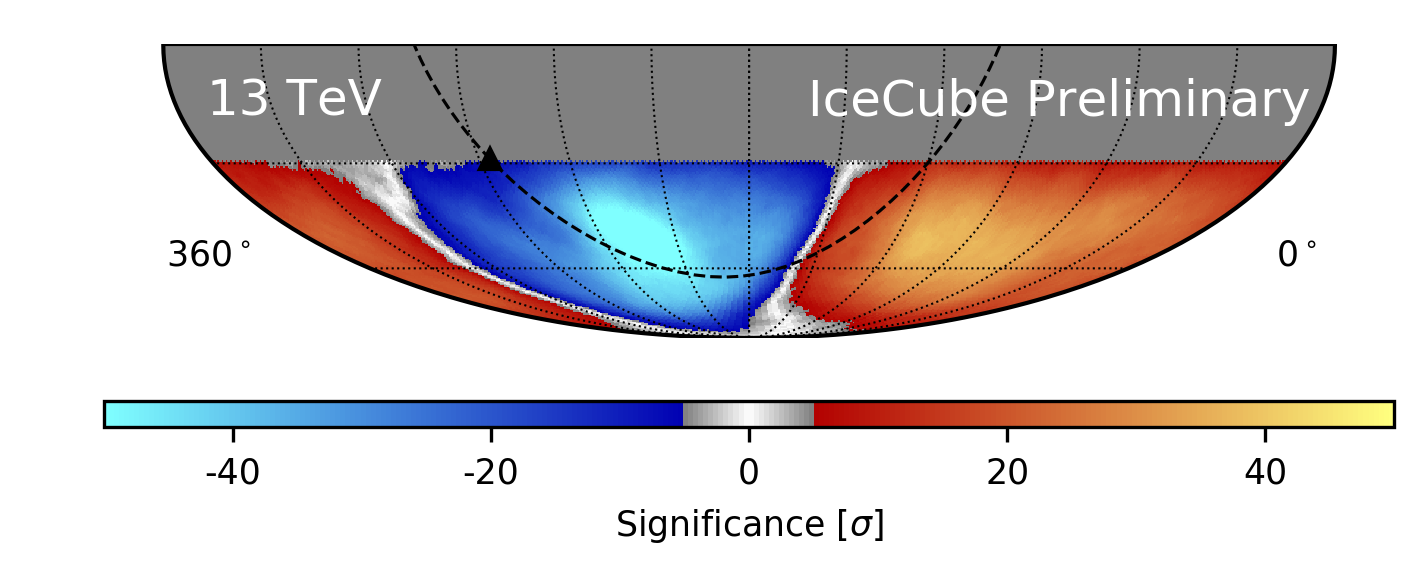}
  \includegraphics[width=0.49\textwidth]{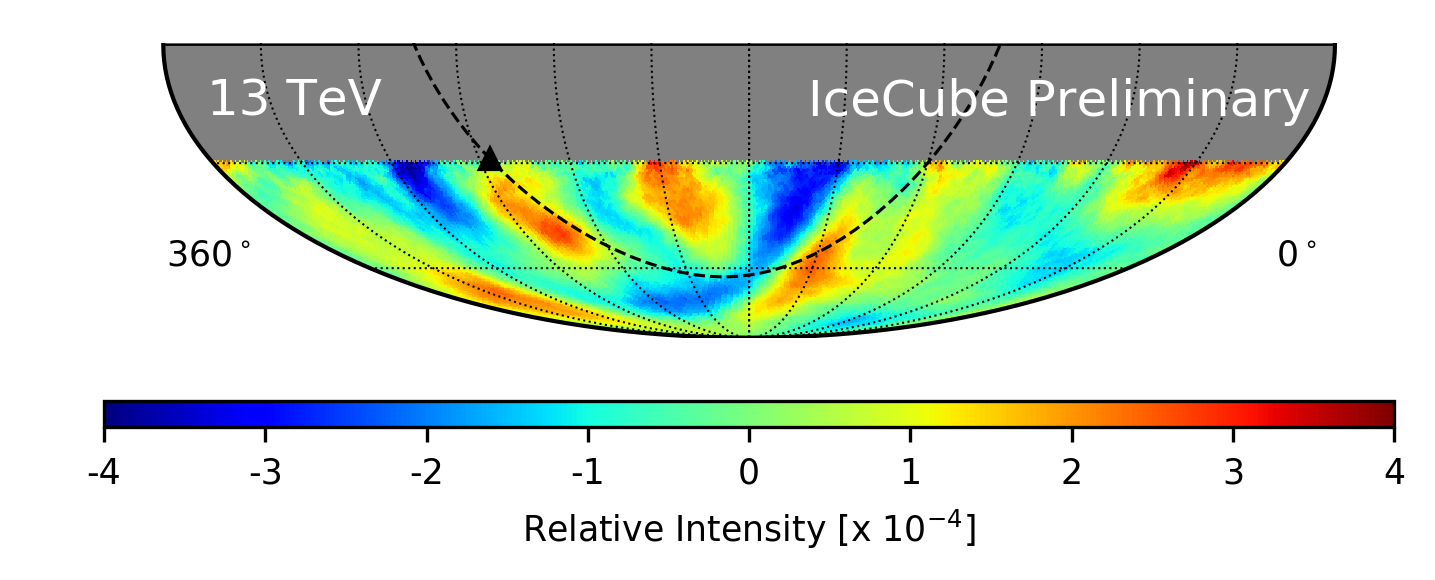}
  \includegraphics[width=0.49\textwidth]{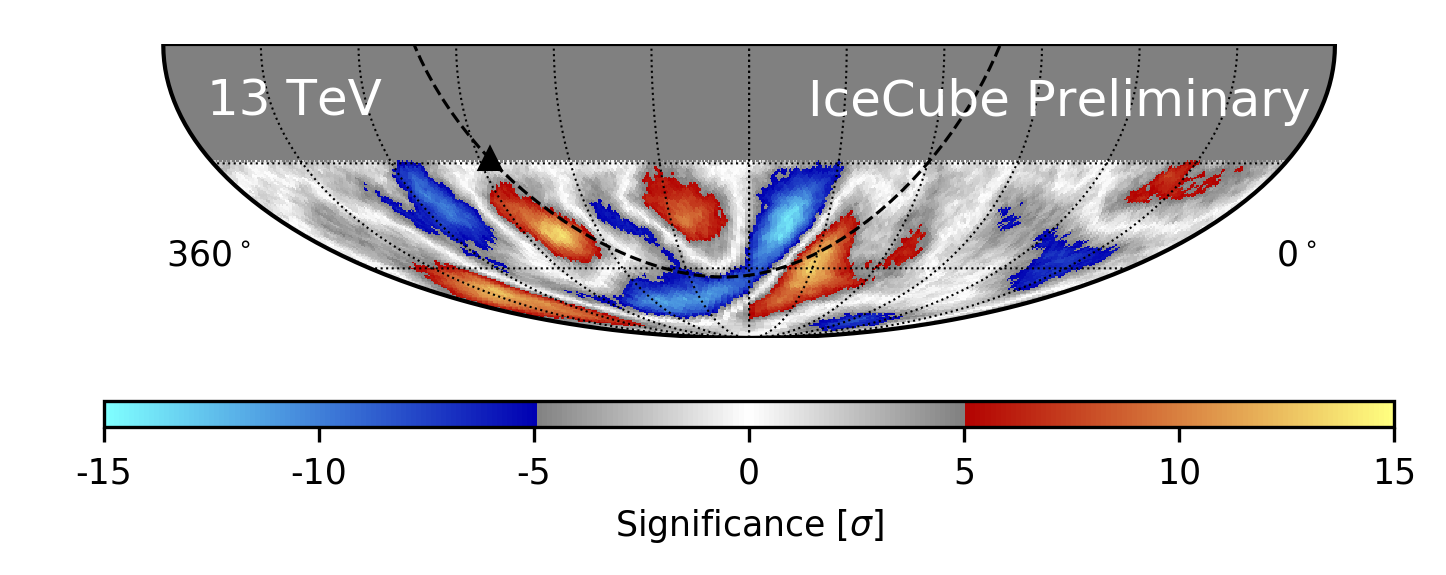}
  \caption{Relative intensity (\textit{left}) and significance (\textit{right}) for events with $\log_{10}(E_\mathrm{reco}/\mathrm{GeV}) < 4.25$. Maps are in equatorial coordinates and have a top-hat smoothing radius of $5^\circ$. The large-scale structure (\textit{top}) is the result of considering all events. The small-scale structure (\textit{bottom}) represents the residual signal after subtracting off the best-fit dipole and quadrupole components. Pixels below a $5\sigma$ threshold in the significance map are shown in grayscale.}
  \label{fig:low_energy_structure}
\end{figure*}

% Large- and small-scale structure
Figure~\ref{fig:low_energy_structure} shows the large- and small-scale structure for the lowest reconstructed energy bin. The large-scale structure map is largely equivalent to the energy-binned maps shown in previous analyses~\citep{aartsen_2016, Abbasi:2021kjd}, with large regions of excess and deficit observed at high significance. The map of small-scale structure is new, having previously only been shown for all events. In previous studies, there was a deficit in the small-scale map located between right ascensions of $60-90^\circ$. No significant structure is present in the low-energy small-scale structure at that location, indicating that previous observations featured signal contamination from high-energy events. 

\begin{figure*}[ht]
  \centering
  \includegraphics[width=0.49\textwidth]{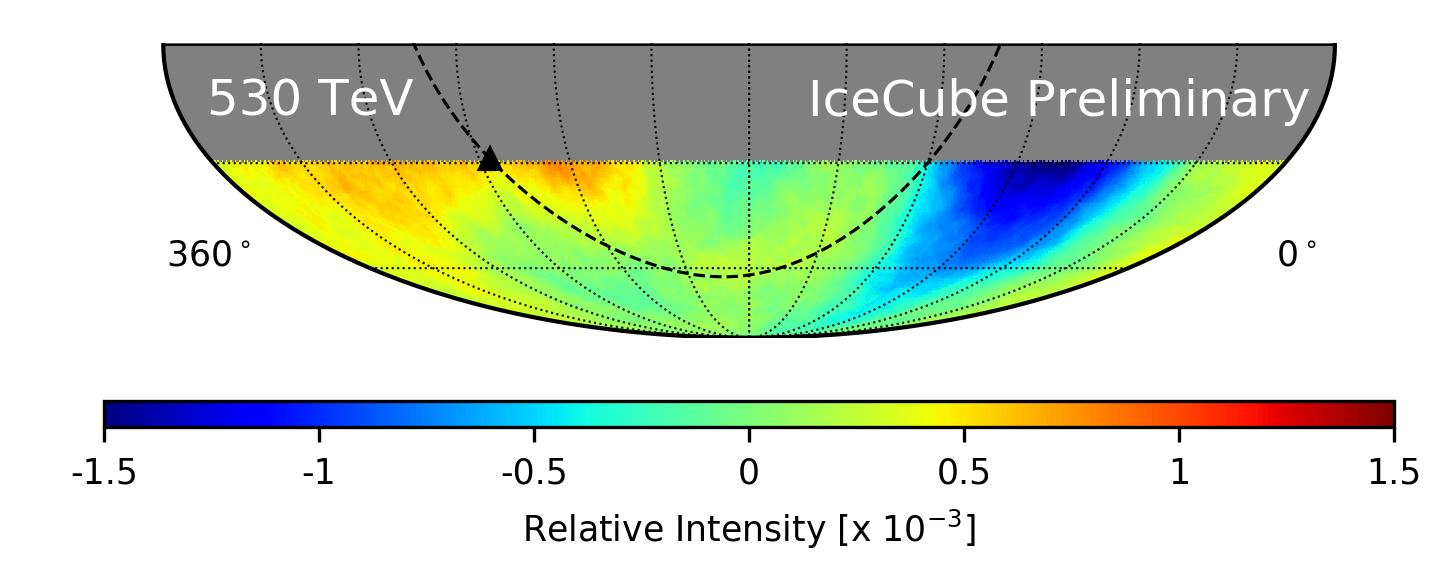}
  \includegraphics[width=0.49\textwidth]{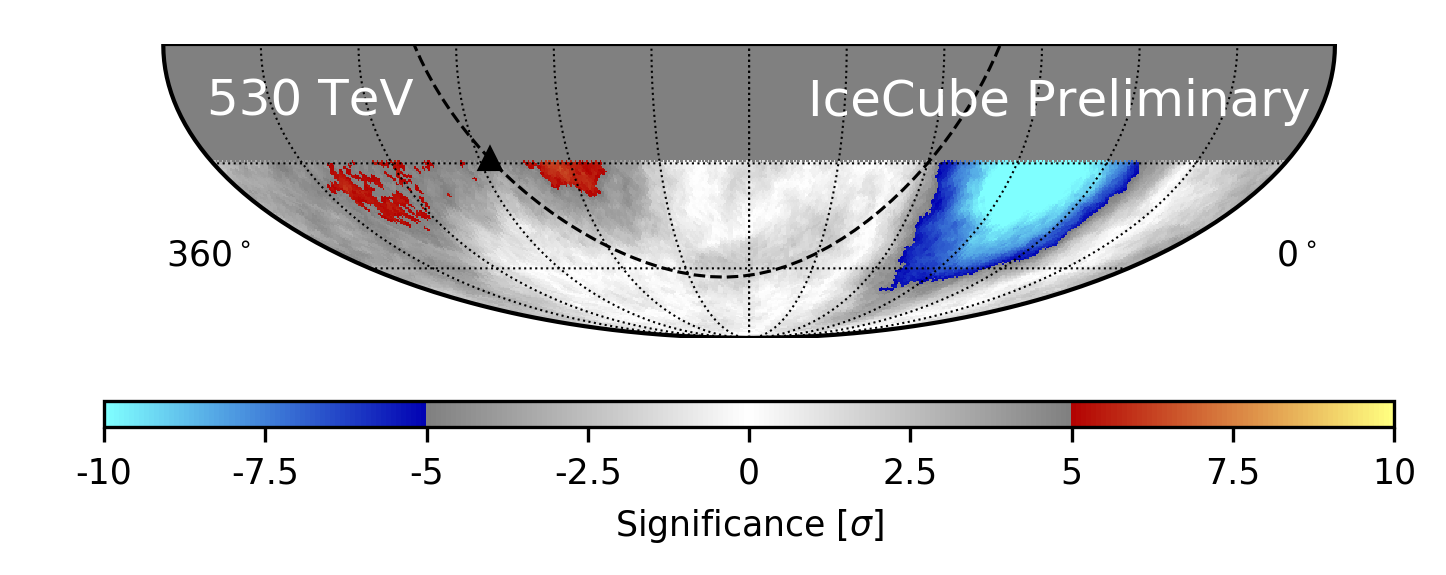}
  \caption{Relative intensity (\textit{left}) and significance (\textit{right}) maps for all events with $\log_{10}(E_\mathrm{reco}/\mathrm{GeV}) > 5.5$. Maps are in equatorial coordinates and have a top-hat smoothing radius of $20^\circ$. Pixels below a $5\sigma$ threshold are shown in grayscale.}
  \label{fig:high_energy_structure}
\end{figure*}

At higher energies, the only significant feature continues to be the deficit located at a right ascension of about $80^\circ$, as shown in Fig.~\ref{fig:high_energy_structure}. To account for the reduced statistics at these energies, we used a larger smoothing radius consistent with previous energy analyses ($20^\circ$), as opposed to the $5^\circ$ radius commonly used in large- and small-scale structure studies. The map of the small-scale structure is omitted as, even with this larger smoothing radius, no significant structure is observed.

\begin{figure*}[ht]
  \centering
  \includegraphics[width=0.49\textwidth]{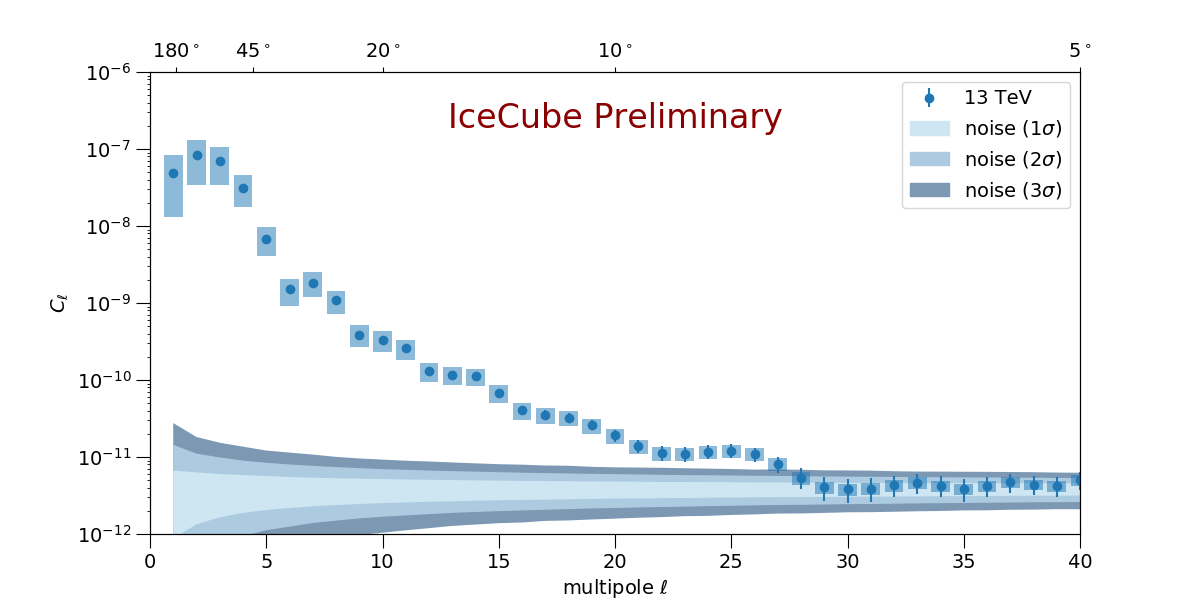}
  \includegraphics[width=0.49\textwidth]{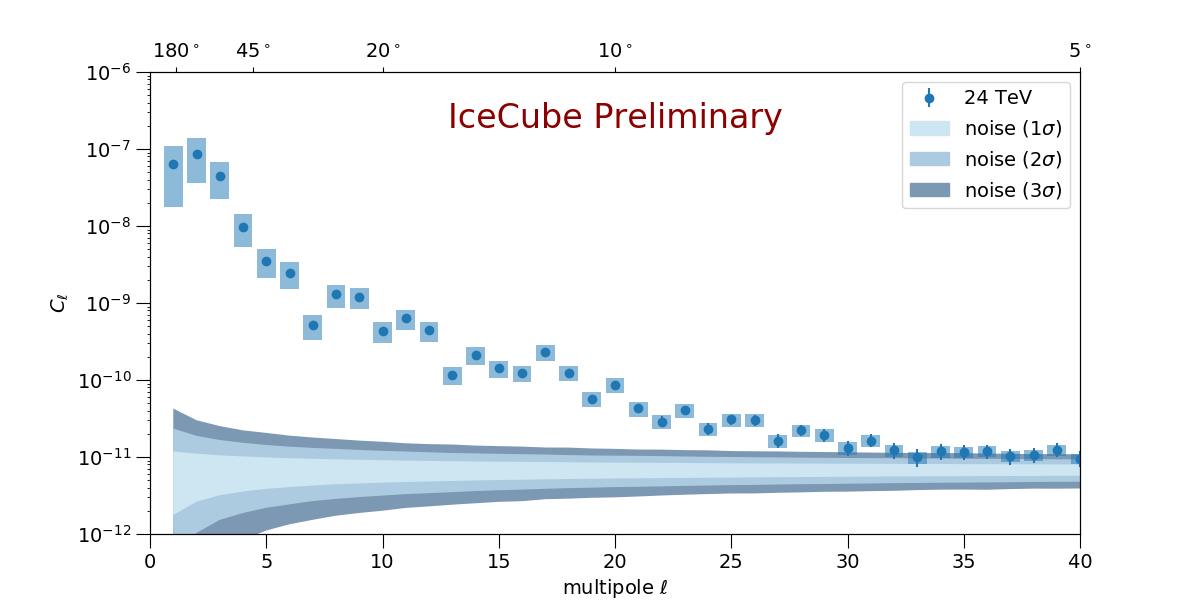}
  \includegraphics[width=0.49\textwidth]{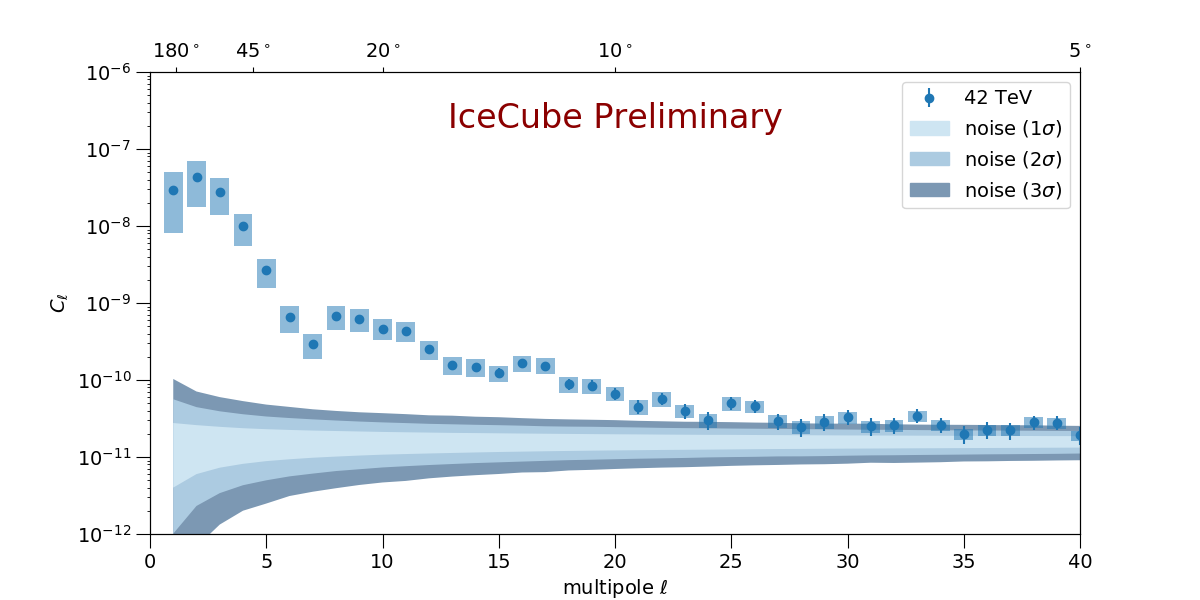}
  \includegraphics[width=0.49\textwidth]{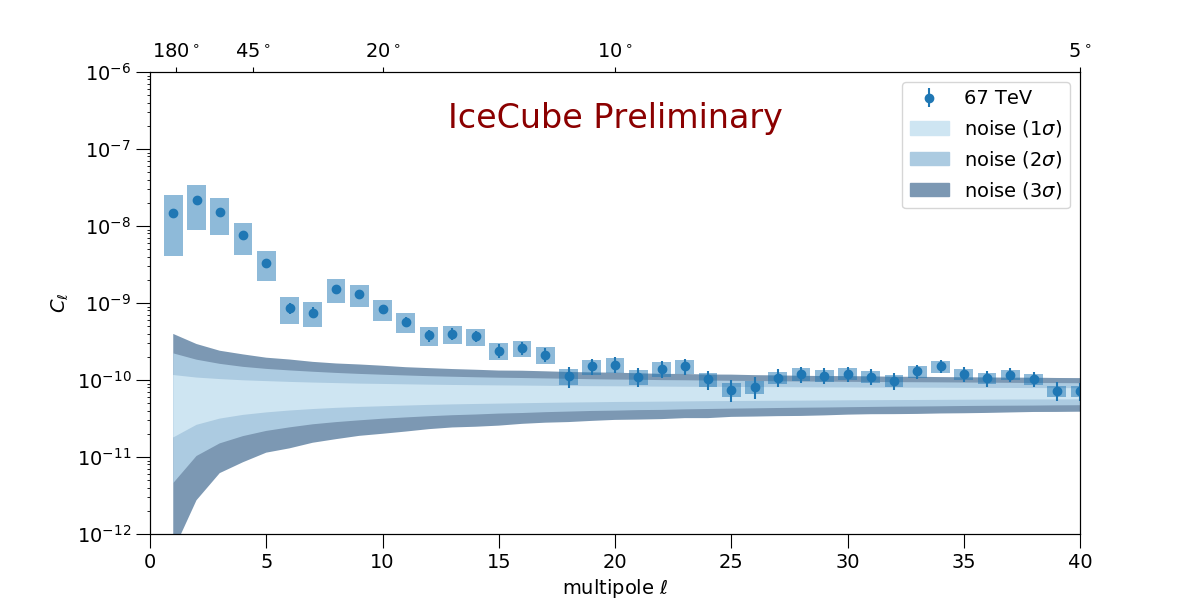}
  \includegraphics[width=0.49\textwidth]{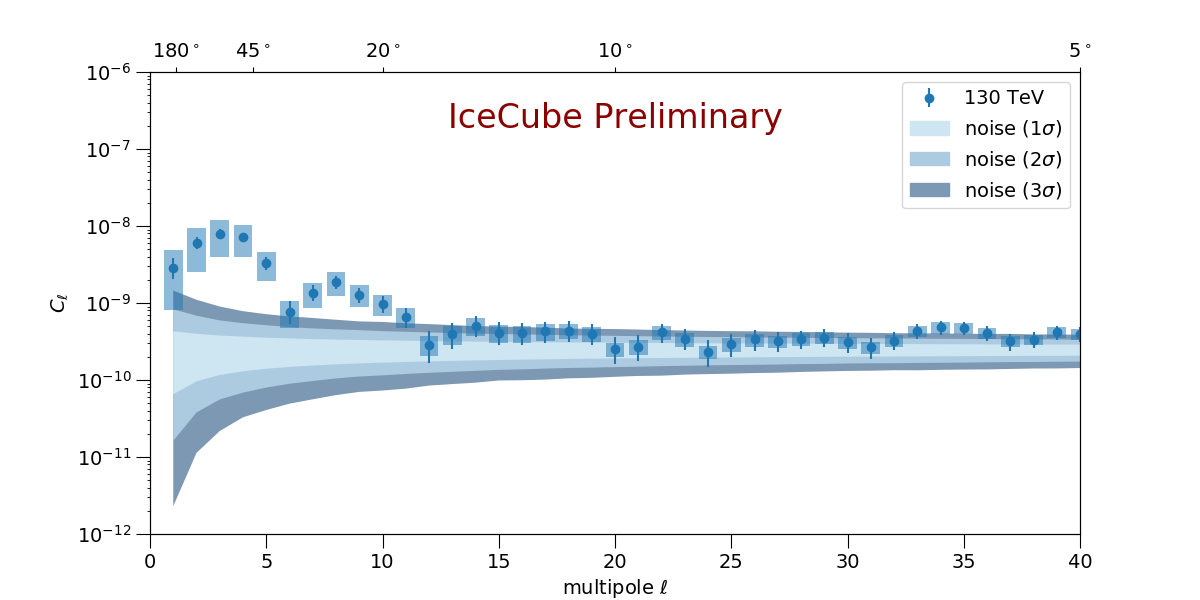}
  \includegraphics[width=0.49\textwidth]{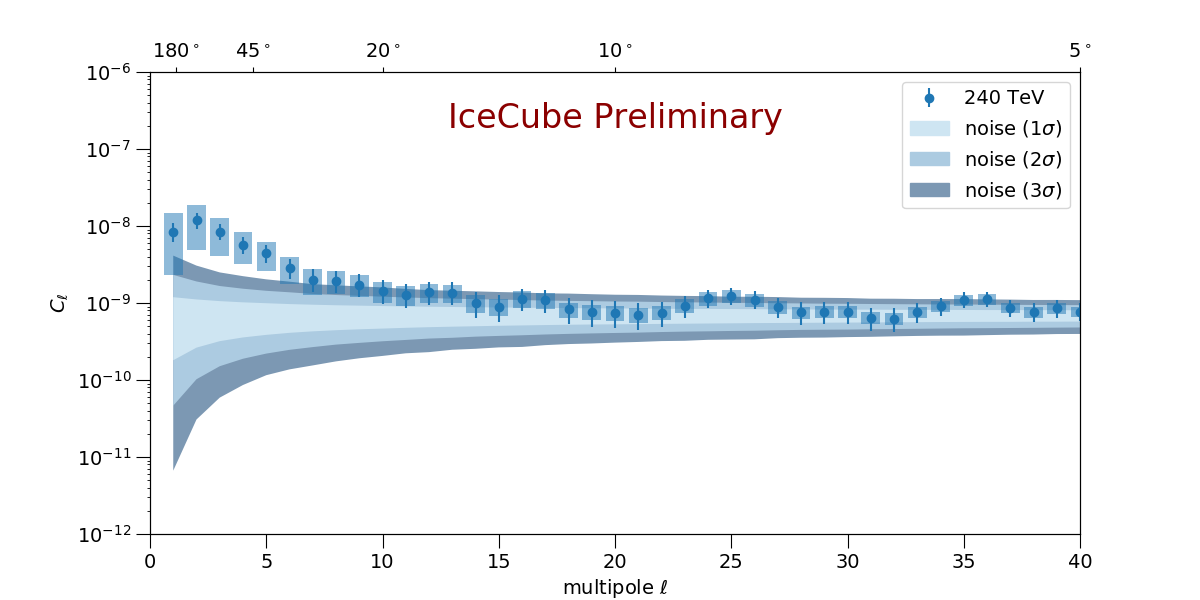}
  \includegraphics[width=0.49\textwidth]{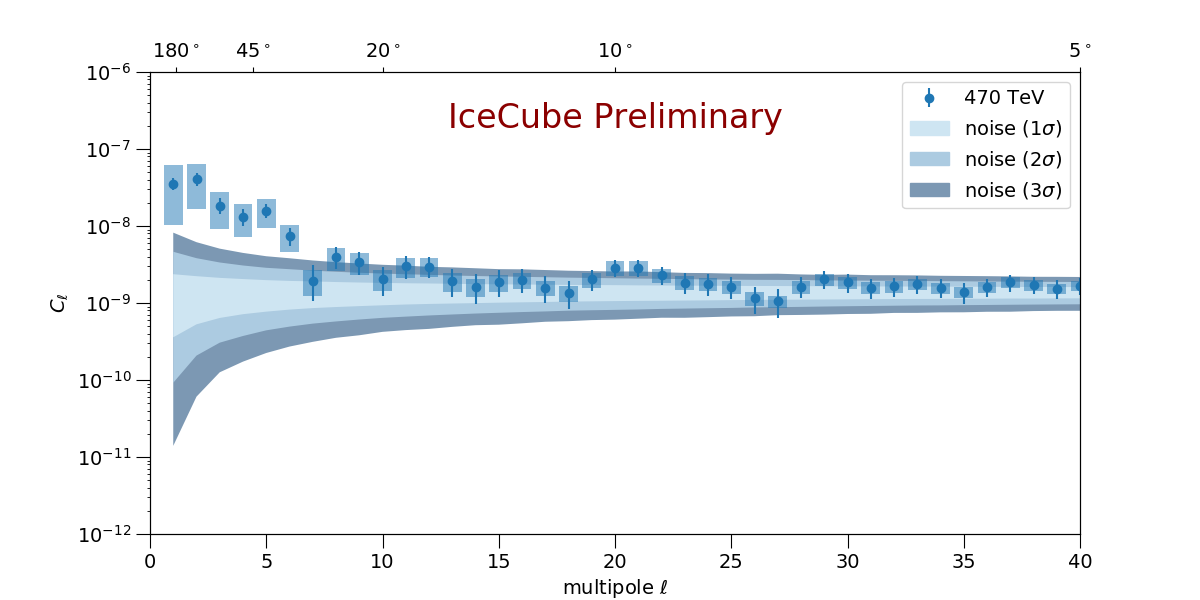}
  \includegraphics[width=0.49\textwidth]{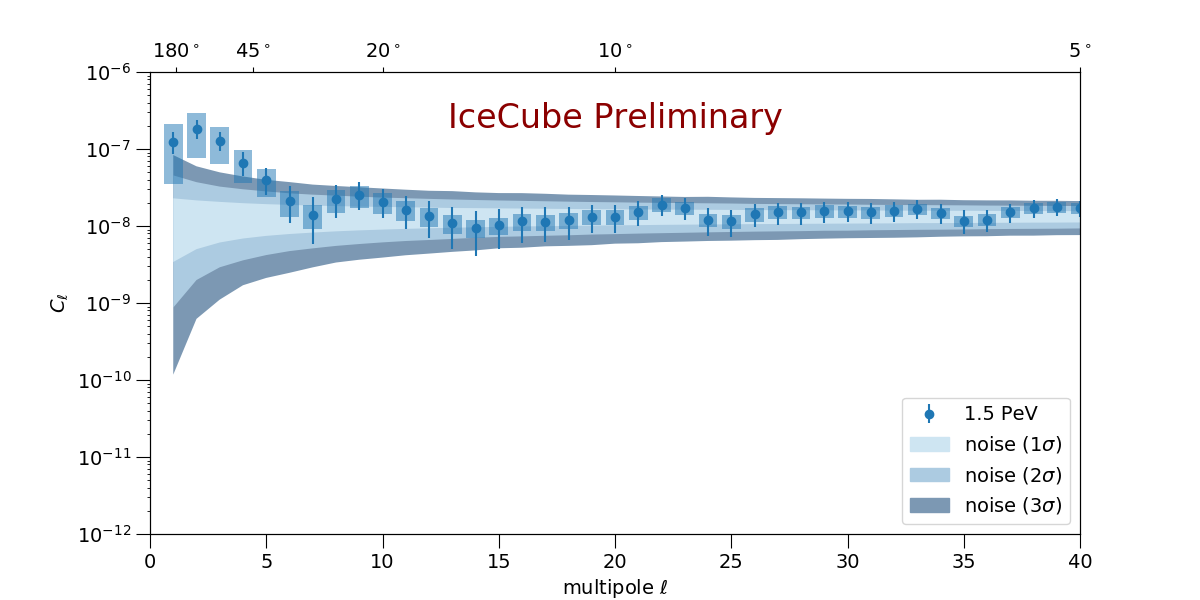}
  \caption{Angular power spectra for the energy bins shown in Fig.~\ref{fig:energy_sim}. Systematic uncertainties (\textit{shaded boxes}) represent a standard deviation of power spectra calculated using identical input $C_\ell$ values with random orientations. Statistical uncertainties (\textit{error bars}) are calculated by fluctuating pixel counts within Poisson uncertainties. The large shaded bands represent the reconstructed power from an isotropic sky. The highest energy bin ($\log_{10}(E_\mathrm{reco}/\mathrm{GeV}) > 6.5$) is omitted, as its power lies within the isotropic bands at all multipoles.}
  \label{fig:powerspec}
\end{figure*}

% Power spectra split by energy
The increased statistics of the 11-year dataset allow for IceCube's first look at the angular power spectrum of the cosmic ray anisotropy as a function of energy. Figure~\ref{fig:powerspec} shows the power spectra for each energy bin, excepting the highest-energy one where we lack sufficient statistics. The shaded noise bands represent 1, 2, and $3\sigma$ containment for power spectra produced in response to an isotropic sky, simulated by varying background maps within Poisson uncertainties. The increasing power of the noise floor with energy is the result of decreasing statistics. It is important to note that these plots display the \textit{observed} power spectra --- IceCube's partial sky coverage results in correlation between multipoles, producing a ringing effect. This effect is avoidable only through collaborative full-sky analyses with other detectors~\citep{Abeysekara_2019}.

\begin{figure}[ht]
  \centering
  \includegraphics[width=\textwidth]{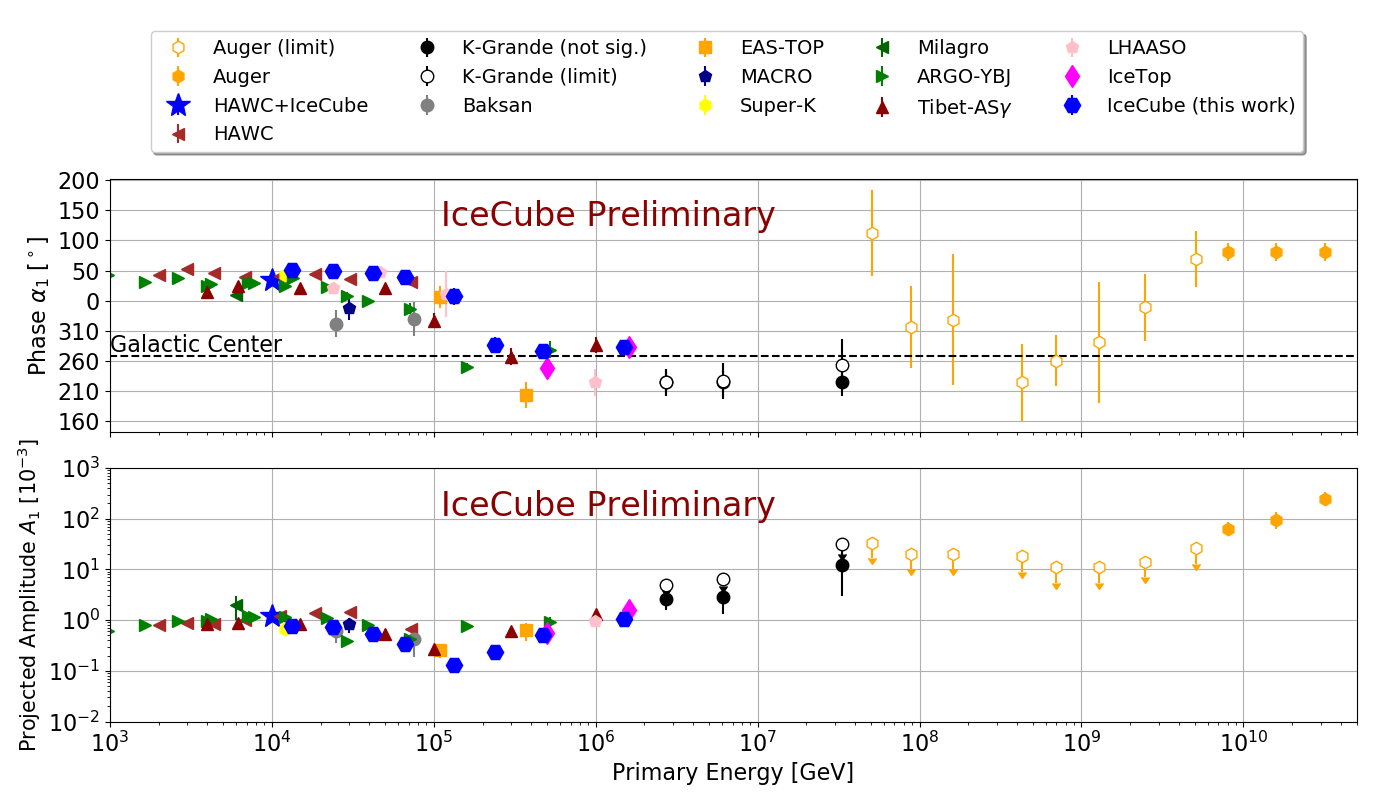}
  \caption{Phase (\textit{top}) and amplitude (\textit{bottom}) of the best-fit dipole component as a function of reconstructed energy, shown in comparison to results from other experiments. The phase and amplitude are calculated using a harmonic fit up to $\ell = 3$ for a one-dimensional projection of the relative intensity along right ascension. Results shown are from~\cite{
abeysekara_2018, Apel_2019, ALEKSEENKO2009179, Aglietta:2009mu, PhysRevD.67.042002, guillian_2007, Abdo:2008aw, Bartoli_2015, bartoli2018, Amenomori_2005, Aartsen:2012ma, Abeysekara_2019, lhaaso_2021, Auger2020ApJ}.}
  \label{fig:dipole}
\end{figure}

% Dipole phase and amplitude
An alternative method of visualizing the change in the angular power spectra over energy is to plot the behavior of a single multipole. Figure~\ref{fig:dipole} shows the phase and amplitude of the dipole component as a function of energy, as compared to the results from other experiments. The calculation methods for this plot are slightly different; following the procedure used in Ref.~\citep{aartsen_2016}, we instead use a harmonic fit (up to $\ell = 3$) on a one-dimensional projection of the relative intensity as a function of right ascension. The resultant dipole component displays a transition in phase and amplitude occurring near 100\,TeV --- consistent with previous work in IceCube and results from other experiments.

\begin{figure}[ht]
  \centering
  \includegraphics[width=\textwidth]{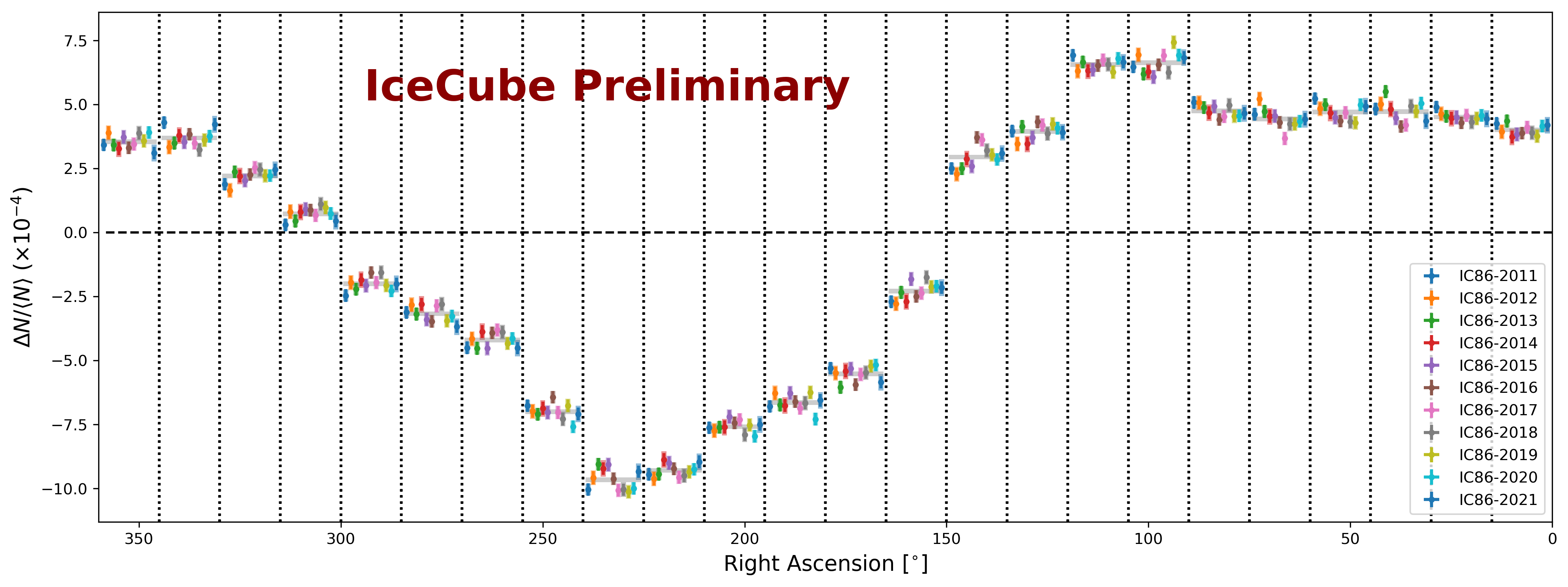}
  \caption{Comparison of relative intensity as a function of time, shown as a one-dimensional projection binned in right ascension. Each series represents one calendar year, beginning and ending in May. Both statistical (\textit{error bars}) and systematic (\textit{shaded boxes}) uncertainties are included, the latter being calculated from the root-mean-square of the anti-sidereal distribution for the corresponding year. The narrow gray band in each bin indicates the relative intensity and uncertainty for the combined dataset.}
  \label{fig:proj_1d}
\end{figure}

% Time-dependence of signal
In order to look for time-dependent variation in the anisotropy, we use the same one-dimensional projection as when finding the phase and amplitude of a best-fit dipole. Figure~\ref{fig:proj_1d} shows the relative intensity for bins in right ascension, with each data series representing a different calendar year. Use of calendar years greatly reduces the systematic uncertainty of solar influences on the sidereal signal as seen in the anti-sidereal frame, making statistical and systematic uncertainties of similar magnitude~\citep{Abbasi:2021kjd}. Combined with the extended observation period, the reduced systematic uncertainties hint at the potential for time-dependent behavior in some bins, where the variations are greater than expected systematic variations about the 11-year average values shown in gray. A study dedicated to determining the significance of this effect is planned.

\section{Conclusion}\label{sec3}

%The data-taking period covers a complete solar cycle, providing new insight into the time variability of the signal. We present preliminary results using this up-to-date event sample.

We have analyzed over 700 billion cosmic-ray-induced muon events collected by the IceCube Neutrino Observatory between May 2011 and May 2022 to study the arrival direction distribution of cosmic rays in the TeV-PeV energy range. The increased event volume by a factor of two since our last report~\cite{aartsen_2016} made it possible to improve the significance of the large- and small-scale structures of the cosmic ray anisotropy up to PeV energies and down to scales of $6^\circ$. 

The latest study includes several improvements to previous IceCube analyses of the observed anisotropy in cosmic ray arrival direction. The observed angular structures and power spectrum are consistent with previous measurements, but we can now report significant features in our highest-energy map. In addition, we observe a variation in the angular power spectrum as a function of energy, hinting at a relative decrease in large-scale features above 100\,TeV.

In addition to greater statistics, the use of single detector geometry enables the study of data by calendar year as opposed to detector year and thus reduces systematic uncertainties in the sidereal anisotropy that arise from interference with the solar frame. Additional improvements in our simulations and a larger Monte Carlo sample result in better energy reconstructions.

%\section{Listing some References}\label{sec2}

%As we discussed in \textsection\ref{sec1}, etc. This is a paper from a previous ICRC \cite{Zoll:2015wcu}. This is a second paper from a previous ICRC \cite{Peiffer:2017vsm}. This is a paper from the current ICRC \cite{Author:2023icrc}.
%Here is an IceCube journal paper \cite{Aartsen:2016nxy} and an external journal paper \cite{Waxman:1998yy}.

% Bibtex references:
\bibliographystyle{ICRC}
{\footnotesize
\bibliography{references}
}
% Alternatively, you can include references by hand:
%\begin{thebibliography}{99}
%\bibitem{...}
%
%\end{thebibliography}

\clearpage

%The following list of authors, affiliations and funding agencies will be updated at the day of submission. The following template is a placeholder generated via https://authorlist.icecube.wisc.edu/icecube on March 25, 2023 and will be updated.
\section*{Full Author List: IceCube Collaboration}

\scriptsize
\noindent
R. Abbasi$^{17}$,
M. Ackermann$^{63}$,
J. Adams$^{18}$,
S. K. Agarwalla$^{40,\: 64}$,
J. A. Aguilar$^{12}$,
M. Ahlers$^{22}$,
J.M. Alameddine$^{23}$,
N. M. Amin$^{44}$,
K. Andeen$^{42}$,
G. Anton$^{26}$
% ...
\\
\\
% ...
$^{12}$ Universit{\'e} Libre de Bruxelles, Science Faculty CP230, B-1050 Brussels, Belgium \\
% ...
$^{17}$ Department of Physics, Loyola University Chicago, Chicago, IL 60660, USA \\
$^{18}$ Dept. of Physics and Astronomy, University of Canterbury, Private Bag 4800, Christchurch, New Zealand \\
% ...
$^{22}$ Niels Bohr Institute, University of Copenhagen, DK-2100 Copenhagen, Denmark \\
$^{23}$ Dept. of Physics, TU Dortmund University, D-44221 Dortmund, Germany \\
% ...
$^{26}$ Erlangen Centre for Astroparticle Physics, Friedrich-Alexander-Universit{\"a}t Erlangen-N{\"u}rnberg, D-91058 Erlangen, Germany \\
% ...
$^{40}$ Dept. of Physics and Wisconsin IceCube Particle Astrophysics Center, University of Wisconsin{\textendash}Madison, Madison, WI 53706, USA \\
% ...
$^{42}$ Department of Physics, Marquette University, Milwaukee, WI, 53201, USA \\
% ...
$^{44}$ Bartol Research Institute and Dept. of Physics and Astronomy, University of Delaware, Newark, DE 19716, USA \\
% ...
$^{63}$ Deutsches Elektronen-Synchrotron DESY, Platanenallee 6, 15738 Zeuthen, Germany  \\
$^{64}$ Institute of Physics, Sachivalaya Marg, Sainik School Post, Bhubaneswar 751005, India \\
% ...

\subsection*{Acknowledgements}

\noindent
The authors gratefully acknowledge the support from the following agencies and institutions:
USA {\textendash} U.S. National Science Foundation-Office of Polar Programs,
U.S. National Science Foundation-Physics Division,
U.S. National Science Foundation-EPSCoR,
Wisconsin Alumni Research Foundation,
Center for High Throughput Computing (CHTC) at the University of Wisconsin{\textendash}Madison,
Open Science Grid (OSG),
Advanced Cyberinfrastructure Coordination Ecosystem: Services {\&} Support (ACCESS),
Frontera computing project at the Texas Advanced Computing Center,
U.S. Department of Energy-National Energy Research Scientific Computing Center,
Particle astrophysics research computing center at the University of Maryland,
Institute for Cyber-Enabled Research at Michigan State University,
and Astroparticle physics computational facility at Marquette University;
Belgium {\textendash} Funds for Scientific Research (FRS-FNRS and FWO),
FWO Odysseus and Big Science programmes,
and Belgian Federal Science Policy Office (Belspo);
Germany {\textendash} Bundesministerium f{\"u}r Bildung und Forschung (BMBF),
Deutsche Forschungsgemeinschaft (DFG),
Helmholtz Alliance for Astroparticle Physics (HAP),
Initiative and Networking Fund of the Helmholtz Association,
Deutsches Elektronen Synchrotron (DESY),
and High Performance Computing cluster of the RWTH Aachen;
Sweden {\textendash} Swedish Research Council,
Swedish Polar Research Secretariat,
Swedish National Infrastructure for Computing (SNIC),
and Knut and Alice Wallenberg Foundation;
European Union {\textendash} EGI Advanced Computing for research;
Australia {\textendash} Australian Research Council;
Canada {\textendash} Natural Sciences and Engineering Research Council of Canada,
Calcul Qu{\'e}bec, Compute Ontario, Canada Foundation for Innovation, WestGrid, and Compute Canada;
Denmark {\textendash} Villum Fonden, Carlsberg Foundation, and European Commission;
New Zealand {\textendash} Marsden Fund;
Japan {\textendash} Japan Society for Promotion of Science (JSPS)
and Institute for Global Prominent Research (IGPR) of Chiba University;
Korea {\textendash} National Research Foundation of Korea (NRF);
Switzerland {\textendash} Swiss National Science Foundation (SNSF);
United Kingdom {\textendash} Department of Physics, University of Oxford.

\end{document}